\documentclass[12pt]{article}

\usepackage{amssymb,amsmath} %for Bbb, blacktriangleright etc

\newtheorem{theorem}{Theorem}[section]
\newtheorem{lemma}{Lemma}[section]
\newtheorem{corollary}{Corollary}[section]

\newtheorem{remark}{Remark}[section]
\newtheorem{question}{Question}[section]

\newcommand{\bull}{\mbox{$\;\;\;$\vrule height .9ex width .8ex depth -.1ex}}
\newcommand{\qed}{$\;\Box$}
\newenvironment{proof}{\par\smallbreak\noindent{\bf Proof.~}}
{\unskip\nobreak\hfill \bull \par\medbreak}
\newenvironment{claim}{\par\smallskip\noindent{\it Claim.}}{\par}
\newenvironment{proofofclaim}{\par\smallbreak\noindent{\it Proof of Claim.}}%
{\qed \par\smallbreak}

\newcommand{\dlogtime}{\mbox{\rm DLOGTIME}}

\newcommand{\function}[2]{:#1 \rightarrow #2}

\renewcommand{\L}{\mbox{\rm L}}
\newcommand{\NL}{\mbox{\rm NL}}
\newcommand{\tc}[1]{\mbox{\rm TC$^{#1}$}}
\newcommand{\nc}[1]{\mbox{\rm NC$^{#1}$}}
\newcommand{\ac}[1]{\mbox{\rm AC$^{#1}$}}
\newcommand{\aut}{\mathop{\rm Aut}\nolimits}

\newcommand{\rig}[1]{\mathit{rig}(#1)}
\newcommand{\WL}[1]{\mbox{\rm WL$^{#1}$}}

\title{From Invariants to Canonization in Parallel}

\author{
Johannes K\"obler%
\thanks{Institut f\"ur Informatik,
Humboldt Universit\"at zu Berlin, D-10099 Berlin}%
\ \ and\ \ Oleg Verbitsky%
\thanks{IAPMM, 79060 Lviv, Ukraine.
Supported by an Alexander von Humboldt fellowship.}
}

\date{}

\begin{document}

\sloppy

\maketitle

\begin{abstract}
  A function $f$ of a graph is called a \emph{complete graph
    invariant} if two given graphs $G$ and $H$ are isomorphic exactly
  when $f(G)=f(H)$. If additionally, $f(G)$ is a graph isomorphic to
  $G$, then $f$ is called a \emph{canonical form} for graphs.
  Gurevich \cite{Gur97} proves that any polynomial-time computable
  complete invariant can be transformed into a polynomial-time
  computable canonical form.  We extend this equivalence to the
  polylogarithmic-time model of parallel computation for classes of
  graphs having either bounded rigidity index or small separators. In
  particular, our results apply to three representative classes of
  graphs embeddable into a fixed surface, namely, to 3-connected
  graphs admitting either a polyhedral or a large-edge-width embedding
  as well as to all embeddable 5-connected graphs.  Another
  application covers graphs with treewidth bounded by a constant $k$. Since for the
  latter class of graphs a complete invariant is computable in \nc{}, it
  follows that graphs of bounded treewidth have a canonical form
  (and even a canonical labeling) computable in~\nc{}.
\end{abstract}

\maketitle

\section{Introduction}

\noindent
We write $G\cong H$ to indicate that $G$ and $H$ are isomorphic
graphs.  A \emph{complete invariant} is a function $f$ on graphs such
that $f(G)=f(H)$ if and only if $G\cong H$.  If, in addition, $f(G)$
is a graph isomorphic to $G$, then $f$ is called a \emph{canonical
  form} for graphs.  For a given graph $G$ and a one-to-one map
$\sigma$ on the vertices of $G$, we use $G^\sigma$ to denote the
isomorphic image of $G$ under $\sigma$.  A \emph{canonical labeling}
assigns to each graph $G$ a map $\sigma$ so that the function $f$
defined as $f(G)=G^\sigma$ is a complete invariant.  Note that $f$ is
even a canonical form.  Thus, the notion of a canonical labeling is
formally stronger than that of a canonical form which in turn is
formally stronger than that of a complete invariant.

Obviously, a polynomial-time computable complete invariant can be used
to decide in polynomial time whether two given graphs are isomorphic.
Conversely, it is not known whether a polynomial-time decision
algorithm for graph isomorphism implies the existence of a
polynomial-time complete invariant (cf.\ the discussion in
\cite[Section 5]{AT05}).  However, for many classes of graphs
for which we have an efficient isomorphism test, we also have a
canonical labeling algorithm of comparable complexity (see, e.g.,
\cite{MR91,Lin92,Bus97}); but this often requires substantial
additional efforts (cf., e.g., \cite{BL83,ADM06}).

Gurevich \cite{Gur97} proves that a polynomial-time computable
complete graph invariant can be used to compute a canonical labeling
in polynomial time.  This result is really enlightening because there
are approaches to the graph isomorphism problem which are based on
computing a graph invariant and, without an extra work, do not provide
us with a canonical form. An important example is the
\emph{$k$-dimensional Weisfeiler-Lehman algorithm} $\WL{k}$.  Given an
input graph $G$, the algorithm outputs a coloring of its vertices in
polynomial time (where the degree of the polynomial bounding the
running time depends on $k$).  $\WL{k}$ always produces the same
output for isomorphic input graphs.  Whether the algorithm is able to
distinguish $G$ from every non-isomorphic input graph $H$ depends on
whether the dimension $k$ is chosen large enough for $G$.  In
particular, $k=1$ suffices for all trees $T$.  However, notice that
the coloring computed by $\WL{1}$ on input $T$ partitions the vertex
set of $T$ into the orbits of the automorphism group of $T$ and hence
$\WL{1}$ does not provide a canonical labeling unless $T$ is rigid
(i.e., $T$ has only the trivial automorphism). We mention that an
appropriate modification of the 1-dimensional Weisfeiler-Lehman
algorithm to a canonical labeling algorithm is suggested
in~\cite{IL90a}.

The reduction of a canonical labeling to a complete invariant presented in \cite{Gur97} 
(as well as in \cite{IL90a}) is inherently sequential
and thus leaves open the following question. 

\begin{question}\label{que:nc}
Suppose that for the graphs in a certain class $C$
we are able to compute a complete invariant in \nc{}. Is it then possible 
to compute also a canonical labeling for these graphs in \nc{}?
\end{question}

\noindent
For several classes of graphs, \nc{} algorithms for computing a complete
invariant are known (see, e.g., \cite{CDR88,MR91,Lin92,GV06}).  For
example, in \cite{GV06} it is shown that a $k$-dimensional
Weisfeiler-Lehman algorithm making logarithmically many rounds can be
implemented in $\tc1\subseteq\nc2$ and that such an algorithm succeeds
for graphs of bounded treewidth.  Similar techniques apply also to
planar graphs but for this class a canonical labeling algorithm in
\ac1 is known from an earlier work \cite{MR91}. Nevertheless, also in
this case it is an interesting question whether the approach to the
planar graph isomorphism problem suggested in \cite{GV06}, which is
different from the approach of \cite{MR91}, can be adapted for finding
a canonical labeling. Finally, Question \ref{que:nc} even makes sense
for classes $C$ for which we don't know of any \nc{}-computable
complete invariant since such an invariant may be found in the future.

We notice that a positive answer to Question~\ref{que:nc} also implies
that the search problem of computing an isomorphism between two given
graphs in $C$, if it exists, is solvable in \nc{} whenever for $C$ we
have a complete invariant in \nc{} (notice that the known
polynomial-time reduction of this search problem to the decision
version of the graph isomorphism problem is very sequential in nature,
see \cite{KST93}).

As our main result we give an affirmative answer to
Question~\ref{que:nc} for any class of graphs having either small
\emph{separators} (Theorem \ref{thm:sep}) or bounded \emph{rigidity
  index} (Theorem \ref{thm:rig}). A quite general example for a class of
graphs having small separators is the class of graphs whose treewidth is
bounded by a constant. Since, as mentioned above, a complete invariant for these
graphs is computable in \tc1 \cite{GV06}, Theorem \ref{thm:sep}
immediately provides us with an \nc{} (in fact \tc2) canonical
labeling algorithm for such graphs (Corollary \ref{cor:btw}). As a
further application we also get a \tc2 algorithm for solving the
search problem for pairs of graphs in this class (Corollary
\ref{cor:decsearch}).

Regarding the second condition we mention the following representative
classes of graphs with bounded rigidity index:
\begin{itemize}
\item The class of 3-connected graphs having a \emph{large-edge-width}
  embedding into a fixed surface $S$ (Corollary \ref{cor:fromTho}).%
  \footnote{ This result is actually stated in a stronger form,
    without referring to a parameter $S$.}
\item The class of 3-connected graphs having a \emph{polyhedral}
  embedding into a fixed surface $S$ (Corollary \ref{cor:fromMRo}).
\item The class of 5-connected graphs embeddable into a fixed surface
  $S$ (Corollary \ref{cor:fromFMo}).
\end{itemize}
As shown by Miller and Reif \cite{MR91}, the canonization problem for
any hereditary class of graphs $C$ (meaning that $C$ is closed under
induced subgraphs) \ac1\ reduces to the canonization problem for the
class of all 3-connected graphs in $C$. Thus, with respect to the
canonization problem, the 3-connected case is of major interest.

The rest of the paper is organized as follows. In Section \ref{s:defn}
we provide the necessary notions and fix notation.  Graphs with small
separators are considered in Section \ref{s:sep} and graphs with
bounded rigidity index are considered in Section \ref{s:rig}.  Section
\ref{s:concl} summarizes our results and discusses remaining open
problems.

\section{Preliminaries}\label{s:defn}

\noindent
We assume familiarity with basic notions of complexity theory such as
can be found in the standard books in the area.
In particular, we simply recall the NC hierarchy of
polylogarithmic-time parallel complexity classes. Namely,
$\nc{}=\textstyle\bigcup_{i\ge1}\nc i$, where \nc i\ consists of
functions computable by \dlogtime-constructible boolean circuits of polynomial size
and depth $O(\log^in)$.
%with internal $\mathit{AND}$, $\mathit{OR}$ and $\mathit{NOT}$
%gates with bounded fan-in. 
The class \ac i\ is the extension of \nc i\ to circuits with unbounded fan-in
%$\mathit{AND}$, $\mathit{OR}$ and $\mathit{NOT}$ gates 
and \tc i\ is a further extension allowing threshold gates as well.  
Recall also that
$$\ac0\subseteq \tc0\subseteq \nc1 \subseteq \L\subseteq \NL
\subseteq\ac1\mbox{ and }\nc i\subseteq\ac i\subseteq\tc
i\subseteq\nc{i+1},$$
where \L\ (resp.\ \NL) is the set of languages
accepted by (non)deterministic Turing machines using logarithmic
space.  Alternatively, the \ac i\ level of the \nc{} hierarchy can be
characterized as the class of all functions computable by a CRCW PRAM
with polynomially many processors in time $O(\log^in)$.

The vertex set of a graph $G$ is denoted by $V(G)$. The set of all vertices
adjacent to a vertex $v\in V(G)$ is called its \emph{neighborhood} of $v$ and 
denoted by~$\Gamma(v)$.

A \emph{colored graph} $G$, besides the binary adjacency relation, has
unary relations $U_1,\ldots,U_n$ defined on $V(G)$.  If a vertex $v$
satisfies $U_i$, we say that $v$ has color $i$. A vertex is allowed to
have more than one color or none. It is supposed that the number of
colors is equal to the number of vertices in a graph, though some of
the color relations may be empty. A colored graph $\langle
G,U_1,\ldots,U_n\rangle$ will be called a \emph{coloring} of the
underlying graph $G$.  An \emph{isomorphism} between colored graphs
must preserve the adjacency relation as well as the color relations.
Thus, different colorings of the same underlying graph need not be
isomorphic.

We consider only classes of graphs that are closed under isomorphism.  For
a given class of graphs $C$ we use $C^*$ to denote the class containing all
colorings of any graph in $C$.

Let $C$ be a class of graphs and let $f$ be a function mapping graphs to
strings over a finite alphabet.  We call $f$ a \emph{complete
  invariant for $C$} if for any pair of graphs $G$ and $H$ in $C$ we
have $G\cong H$ exactly when $f(G)=f(H)$.  A \emph{canonical labeling
  for $C$} assigns to each graph $G$ on $n$ vertices a one-to-one map
$\sigma\function{V(G)}{\{1,\ldots,n\}}$ such that $f(G)=G^\sigma$ is a
complete invariant for $C$.  Note that a complete invariant $f$
originating from a canonical labeling has an advantageous additional
property: $f(G)\ne f(H)$ whenever $G$ is in $C$ and $H$ is not.
Moreover, it provides us with an isomorphism between $G$ and $H$
whenever $f(G)=f(H)$.

The notions of a complete invariant and of a canonical labeling are
easily extensible to colored graphs. In our proofs, extending these
notions to colored graphs will be technically beneficial and, at the
same time, will not restrict the applicability of our results. In fact,
any available complete-invariant algorithm for some class of graphs
$C$ can be easily extended to $C^*$ without increasing the required
computational resources.  In particular, this is true for the
parallelized version of the multi-dimensional Weisfeiler-Lehman
algorithm suggested in~\cite{GV06}.

\section{Small separators}\label{s:sep}

\noindent
For a given graph $G$ and a set $X$ of vertices in $G$, let $G-X$
denote the graph obtained by removing all vertices in $X$ from $G$.  A
set $X$ is called a \emph{separator} if every connected component of
$G-X$ has at most $n/2$ vertices, where $n$ denotes the number of
vertices of $G$.  A class of graphs $C$ is called \emph{hereditary} if
for every $G\in C$, every induced subgraph of $G$ also belongs to~$C$.

\begin{theorem}\label{thm:sep}
  Let $C$ be a hereditary class of graphs such that for a constant
  $r$, every graph $G\in C$ has an $r$-vertex separator. Suppose that
  $C^*$ has a complete invariant $f$ computable in \tc k (resp.\ \ac
  k) for some $k\ge1$.  Then $C$ has a canonical labeling in \tc{k+1}
  (resp.~\ac{k+1}).
\end{theorem}

\begin{proof}
  Having $f$ in our disposal, we design a canonical labeling algorithm
  for $C$.  Let $G$ be an input graph with vertex set
  $V(G)=\{1,\ldots,n\}$ and assume that $G$ has an $r$-vertex
  separator. We describe a recursive algorithm for finding a canonical
  renumbering $\sigma\function{\{1,\ldots,n\}}{\{1,\ldots,n\}}$. In
  the following, the parameter $d$ refers to the recursion depth.
  Initially $d=1$. Further, set $R=2^r+r$.
  
  For a given sequence $s=(v_1,\ldots,v_r)$ of vertices, let $G_s$
  denote the coloring of $G$ in which $v_i$ receives color $(d-1)R+i$.
  
  For each sequence $s=(v_1,\ldots,v_r)$ in parallel we do the
  following.  First of all, we check if the set $\{v_1,\ldots,v_r\}$
  is a separator.  We are able to do this in \ac1 since checking if
  two vertices are in the same connected component can be done in
  logarithmic space \cite{Rei05} and the remaining job can be easily
  organized in \tc0. If the verification is positive, we mark the
  sequence $s$ as \emph{separating}. If no such sequence $s$ is
  separating, i.e., $G$ has no $r$-vertex separator, we terminate and
  output the identity permutation.  Otherwise, for each separating
  sequence $s$ in parallel we compute $f(G_s)$. Then in \ac1 we find a
  sequence $s=(v_1,\ldots,v_r)$ for which the value $f(G_s)$ is
  lexicographically minimum.
  
  At this stage we are able to determine the renumbering $\sigma$ only
  in a few points.  Namely, we set $\sigma(v_i)=(d-1)R+i$ for each
  $i\le r$.
  
  To proceed further, let $F_1,\ldots,F_m$ be the connected components
  of $G-X$ where $X=\{v_1,\ldots,v_r\}$.  We color each $v\notin X$ by
  its adjacency pattern to $X$, that is, by the set of all neighbors
  of $v$ in $X$, encoding this set by a number in the range between
  $(d-1)R+r+1$ and $dR$. Each $F_j$, regarded as a colored graph, will
  be called an \emph{$X$-flap}. For each $X$-flap $F_j$ in parallel,
  we now compute $f(F_j)$ and establish the lexicographic order
  between these values.  At this stage we fix the following partial
  information about the renumbering $\sigma$ under construction:
  $\sigma(u)<\sigma(v)$ whenever we have $f(F_l)<f(F_j)$ for the two
  flaps $F_l$ and $F_j$ containing $u$ and $v$, respectively.  Thus,
  we split $V(G)\setminus X$ into blocks $V(F_1),\ldots,V(F_m)$ and
  determine the renumbering $\sigma$ first between the blocks.  It may
  happen that for some flaps we have $f(F_l)=f(F_j)$.  We fix the
  $\sigma$-order between the corresponding blocks arbitrarily.  Note
  that the output will not depend on a particular choice made at this
  point.
  
  It remains to determine $\sigma$ inside each block $V(F_j)$.  We do
  this in parallel. For $F_j$ with more than $r$ vertices we repeat
  the same procedure as above with the value of $d$ increased by 1.
  If $F=F_j$ has $t\le r$ vertices, we proceed as follows.  Let $a$ be
  the largest color present in $F$.  We choose a bijection
  $\tau\function{V(F)}{\{1,\ldots,t\}}$ and define $\sigma$ on $V(F)$
  by $\sigma(u)<\sigma(v)$ if and only if $\tau(u)<\tau(v)$.  To make
  the choice, with each such $\tau$ we associate the colored graph
  $F_\tau$ obtained from $F$ by adding new colors, namely, by coloring
  each $v\in V(F)$ with color $a+\tau(v)$. For each $\tau$ we compute
  $f(F_\tau)$ and finally choose the $\tau$ minimizing $f(F_\tau)$ in
  the lexicographic order. Note that, if the minimum is attained by
  more than one $\tau$, the output will not depend on a particular
  choice.
  
  Finally, we have to estimate the depth of the \tc{} (resp.\ \ac{})
  circuit implementing the described algorithm.
  At the recursive step of depth $d$ we deal
  with graphs having at most $n/2^{d-1}$ vertices. It follows that the
  circuit depth does not exceed
  $\log^k_2n+\log^k_2(n/2)+\log^k_2(n/4)+\cdots+\log^k_2(r)\le\log^{k+1}_2n$.
\end{proof}

\begin{remark}\rm\hfill
  \begin{enumerate}
  \item It is easy to see that the proof of Theorem \ref{thm:sep}
    actually provides an \ac{1} Turing-reduction from the problem of
    computing a canonical labeling for $C$ to the problem of computing
    a complete invariant for $C^*$ where also queries to an
    additional \ac1 oracle are allowed.
  \item Theorem \ref{thm:sep} holds true in a formally stronger form:
    also for the class $C^*$ of colored graphs a canonical labeling is
    computable in \ac{k+1} (resp. \tc{k+1}).  This requires only a
    small change in the proof, namely, the coloring of the graphs
    $G_{s,v}$ should be defined with more care as $G$ could now have
    some precoloring.  The same concerns Theorem \ref{thm:rig} in the
    next section.  Moreover, for many classes $C$, both theorems hold
    true when we replace $C^*$ by $C$ (thus weakening the assumptions
    in the theorems). In fact, this can be proved along the same
    lines, we only have to replace the vertex colors with gadgets
    preserving membership of the graphs in the class $C$.
  \end{enumerate}
\end{remark}

\noindent
It is well known that all graphs of treewidth $t$ have a
$(t+1)$-vertex separator \cite{RS86}.  By \cite{GV06}, this class has
a complete invariant computable in \tc1 and therefore is in the scope
of Theorem~\ref{thm:sep}.

\begin{corollary}\label{cor:btw}
  For each constant $t$, a canonical labeling for graphs of treewidth
  at most $t$ can be computed in~\tc2.
\end{corollary}

\noindent
Theorem \ref{thm:sep} also has relevance to the complexity-theoretic
\emph{decision-versus-search} paradigm.  Let $C$ be a class of graphs.
It is well known (see, e.g., \cite{KST93}) that, if we are able to
test isomorphism of graphs in $C^*$ in polynomial time, we are also
able to find an isomorphism between two given isomorphic graphs in $C$
in polynomial time.  As the standard reduction is very sequential in
nature, it is questionable if this implication stays true in the model
of parallel computation. Nevertheless, a canonical labeling
immediately provides us with an isomorphism between two isomorphic
graphs.

\begin{corollary}\label{cor:decsearch}
  For each constant $t$, an isomorphism between isomorphic graphs of
  treewidth at most $t$ can be computed in~\tc2.
\end{corollary}

\section{Bounded rigidity index}\label{s:rig}

\noindent
In this section we show that the canonization problem for any class of
graphs with bounded rigidity index \nc{} reduces to the corresponding
complete invariant problem. Further we show that certain embeddability
properties of a given class of graphs $C$ imply a bound on the rigidity
index of the graphs in $C$.
 
\subsection{Canonizing rigid graphs}

\noindent
A set $S\subseteq V(G)$ of vertices is called \emph{fixing} if every
non-trivial automorphism of $G$ moves at least one vertex in $S$.  The
\emph{rigidity index} of a graph $G$ is defined to be the minimum
cardinality of a fixing set in $G$ and denoted by $\rig G$.

\begin{theorem}\label{thm:rig}
  Let $C$ be a class of graphs such that for a constant $r$, we have
  $\rig G\le r$ for all $G\in C$. Suppose that $C^*$ has a complete
  invariant $f$ computable in \ac k, for some $k\ge1$.  Then $C$ has a
  canonical labeling also in \ac k.
\end{theorem}

\begin{proof}
  Let an input graph $G$ with vertex set $V(G)=\{1,\ldots,n\}$ be given. 
  We describe an algorithm that uses $f$ as a subroutine in
  order to find a canonical renumbering
  $\sigma\function{\{1,\ldots,n\}}{\{1,\ldots,n\}}$ for $G$, provided
  that $G\in C$.
 
  For a given sequence $s=(v_1,\ldots,v_r)$ of vertices, let $G_s$
  denote the coloring of $G$ in which $v_i$ receives color $i$. If $v$
  is another vertex, $G_{s,v}$ denotes the coloring where vertex $v$
  additionally gets color $r+1$.
  
  For each such sequence $s$ in parallel we do the following.  For
  each $v$ in parallel we compute $f(G_{s,v})$.  If all the values
  $f(G_{s,v})$, $v\in V(G)$, are pairwise distinct, which is decidable
  in \ac0, mark $s$ as \emph{fixing}. If no fixing sequence $s$ of
  length $r$ exists, which implies $G\notin C$, we terminate and output the
  identity permutation.  Otherwise, for each fixing sequence $s$ in
  parallel, we compute $f(G_s)$ and determine a sequence
  $s=(v_1,\ldots,v_r)$ for which $f(G_s)$ is lexicographically
  minimum.  For this we use the fact that lexicographic comparison can
  be done in \ac0 and employ a standard \ac1 sorting algorithm.  The
  output permutation $\sigma$ is now computed as follows.  For each
  $i\le r$, we set $\sigma(v_i)=i$.  To determine $\sigma$ everywhere
  else, we sort the values $f(G_{s,v})$ for all $v\in
  V(G)\setminus\{v_1,\ldots,v_r\}$ lexicographically and set
  $\sigma(v)$ to be the number of $v$ in this order increased by $r$.
\end{proof}

\noindent
Notice that the proof of Theorem \ref{thm:rig} actually provides an
\ac{0} Turing-reduction from the problem of computing a canonical
labeling for $C$ to the problem of computing a complete invariant for
$C^*$ where also queries to an additional \tc0 oracle are allowed.

\subsection{Basics of topological graph theory}

\noindent
A detailed exposition of the concepts discussed in this section can be
found in \cite{MT01} (see also \cite[Chapter 7]{GY04}).  We are
interested in embeddability of an abstract graph $G$ into a
surface $S$. 
We only consider undirected graphs without multiple edges and loops. Further, 
all surfaces are supposed to be 2-dimensional,
connected, and closed.

In an \emph{embedding} $\Pi$ of $G$ into $S$, each vertex $v$ of $G$
is represented by a point on $S$ (labeled by $v$ and called
\emph{vertex of the $\Pi$-embedded graph $G$}) and each edge $uv$ of
$G$ is drawn on $S$ as a continuous curve with endpoints $u$ and $v$.
The curves are supposed to be non-self-crossing and any two such
curves either have no common point or share a common endpoint.  A
\emph{face} of $\Pi$ is a connected component of the space obtained
from $S$ by removing the curves.  We consider only \emph{cellular}
embeddings meaning that every face is homeomorphic to an open disc.  A
\emph{closed walk} in a graph is a sequence of vertices $v_1v_2\cdots
v_k$ such that $v_i$ and $v_{i+1}$ are adjacent for any $i<k$,
and $v_1$ and $v_k$ are adjacent as well. Notice that some of
the vertices may coincide.  We will not distinguish between a closed
walk $v_1v_2\cdots v_k$ and any cyclic shift of it or of its reversal
$v_kv_{k-1}\cdots v_1$.  A closed walk $v_1v_2\cdots v_k$ is called
\emph{$\Pi$-facial}, if there exists a face $F$ of $\Pi$, such that
the vertices $v_1,v_2,\ldots, v_k$ occur in this order as labels along
the boundary of $F$.

Two embeddings $\Pi$ and $\Pi'$ of $G$ into $S$ are called
\emph{equivalent} if they can be obtained from each other by a
homeomorphism of $S$ onto itself (respecting vertex labels). Since
such a homeomorphism takes faces of one embedding to faces of the
other embedding, we see that equivalent embeddings have equal sets of
facial walks. In fact, the converse is also true: if the set of the
$\Pi$-facial walks is equal to the set of the $\Pi'$-facial walks,
then $\Pi$ and $\Pi'$ are equivalent. This follows from the fact that
up to homeomorphism, the surface $S$ is reconstructible from the set
of facial walks by attaching an open disc along each facial walk.

A closed walk $v_1v_2\cdots v_k$ can be alternatively thought of as
the sequence of edges $e_1e_2\cdots e_k$ where $e_i=v_iv_{i+1}$
($i<k$) and $e_k=v_1v_k$.  Every edge either appears in two
$\Pi$-facial walks (exactly once in each) or has exactly two
occurrences in a single $\Pi$-facial walk.  An embedding $\Pi$ is
called \emph{polyhedral} if every $\Pi$-facial walk is a cycle (i.e.,
contains at most one occurrence of any vertex) and every two
$\Pi$-facial walks either have at most one vertex in common or share
exactly one edge (and no other vertex).

Let $\aut(G)$ denote the automorphism group of $G$. For a given
automorphism $\alpha\in\aut(G)$, let $\Pi^\alpha$ denote the embedding
of $G$ obtained from $\Pi$ by relabeling the vertices according to
$\alpha$. Note that $\Pi^\alpha$ and $\Pi$ are not necessarily
equivalent (they are \emph{topologically isomorphic}, that is,
obtainable from one another by a surface homeomorphism which is
allowed to ignore the vertex labeling).  An embedding $\Pi$ is called
\emph{faithful} if $\Pi^\alpha$ is equivalent to $\Pi$ for every
automorphism $\alpha\in\aut(G)$.

Recall that a graph $G$ is \emph{$k$-connected} if it has at least
$k+1$ vertices and stays connected after removing any set of at most
$k-1$ vertices.  We now summarize known results showing that, for
$k\ge3$, the flexibility of embedding a $k$-connected graph into
certain surfaces is fairly restricted.
\bigskip

\noindent
{\bf The Whitney Theorem.}~\cite{Whi33} {\it Up to equivalence, every
  3-connected planar graph has a unique embedding into the sphere.}
\bigskip

\noindent
{\bf The Mohar-Robertson Theorem.}~\cite{MR01} {\it Up to equivalence, every connected\,%
\footnote{It is known that only 3-connected graphs have polyhedral embeddings.}  
graph has at most $c_S$ polyhedral embeddings into
a surface $S$, where $c_S$ is a constant depending only on $S$.}
\bigskip

\noindent
A closed curve in a surface is \emph{contractible} if it is homotopic
to a point. The \emph{edge-width} of an embedding $\Pi$ is the minimum length of a 
non-contractible cycle in the $\Pi$-embedded graph.  $\Pi$ is called a 
\emph{large-edge-width embedding} (abbreviated
as \emph{LEW embedding}) if its edge-width is larger than the maximum length 
of a $\Pi$-facial walk. 
\bigskip

\noindent
{\bf The Thomassen Theorem.}~\cite{Tho90} (see also \cite[Corollary
5.1.6]{MT01}) {\it Every 3-connected graph having a LEW embedding 
into a surface $S$ has, up to equivalence, a unique embedding 
into $S$. Moreover, such a surface $S$ is unique.}
\bigskip

\noindent
Note that if a graph has a unique embedding into a surface (as in the Whitney
Theorem or the Thomassen Theorem), then this embedding is faithful.

As we have seen, an embedding is determined by its set of facial walks
(up to equivalence). We will need yet another combinatorial
specification of an embedding. To simplify the current exposition, we
restrict ourselves to the case of orientable surfaces.

Let $G$ be a graph $G$ and let $T$ be a ternary relation on the vertex
set $V(G)$ of $G$. We call $R=\langle G,T\rangle$ a \emph{rotation
  system} of $G$ if $T$ fulfills the following two conditions:

\begin{enumerate}
\item If $T(a,b,c)$ holds, then $b$ and $c$ are in $\Gamma(a)$, the
  neighborhood of $a$ in $G$.
\item For every vertex $a$, the binary relation $T(a,\cdot,\cdot)$ is
  a directed cycle on $\Gamma(a)$ (i.e., for every $b$ there is
  exactly one $c$ such that $T(a,b,c)$, for every $c$ there is exactly
  one $b$ such that $T(a,b,c)$, and the digraph $T(a,\cdot,\cdot)$ is
  connected on $\Gamma(a)$).
\end{enumerate}

\noindent
An embedding $\Pi$ of a graph $G$ into an orientable surface $S$
determines a rotation system $R_\Pi=\langle G,T_\Pi\rangle$ in a
natural geometric way.  Namely, for $a\in V(G)$ and $b,c\in\Gamma(a)$
we set $T_\Pi(a,b,c)=1$ if, looking at the neighborhood of $a$ in the
$\Pi$-embedded graph $G$ from the outside of $S$, $b$ is followed by
$c$ in the clockwise order.

The \emph{conjugate} of a rotation system $R=\langle G,T\rangle$,
denoted by $R^*$, is the rotation system $\langle G,T^*\rangle$, where
$T^*$ is defined as $T^*(a,b,c)=T(a,c,b)$. This notion has two
geometric interpretations.  First, $(R_\Pi)^*$ is a variant of $R_\Pi$
where we look at the $\Pi$-embedded graph from the inside rather than
from the outside of the surface (or, staying outside, just change the
clockwise order to the counter-clockwise order).  Second,
$(R_\Pi)^*=R_{\Pi^*}$ where $\Pi^*$ is a mirror image of $\Pi$.

It can be shown that two embeddings $\Pi$ and $\Pi'$ of $G$ into $S$
are equivalent if and only if $R_\Pi=R_{\Pi'}$ or $R_\Pi=R_{\Pi'}^*$
(see \cite[Corollary 3.2.5]{MT01}).

Further, for a given rotation system $R=\langle G,T\rangle$ and
automorphism $\alpha\in\aut(G)$, we define another rotation system
$R^\alpha=\langle G,T^\alpha\rangle$ by
$T^\alpha(a,b,c)=T(\alpha^{-1}(a),\alpha^{-1}(b),\alpha^{-1}(c))$.  It
is not hard to see that $R^\alpha_\Pi=R_{\Pi^\alpha}$.  If $R=\langle
G,T\rangle$ and $R'=\langle G,T'\rangle$ are two rotation systems of
the same graph $G$ and $R'=R^\alpha$ for some $\alpha\in\aut(G)$, then
this equality means that $\alpha$ is an isomorphism from $R'$ onto $R$
(respecting not only the binary adjacency relation but also the
ternary relations of these structures).

\subsection{Rigidity from non-flexible embeddability}

\noindent
Let $\alpha$ be a mapping defined on a set $V$. We say that $\alpha$
\emph{fixes an element} $x\in V$ if $\alpha(x)=x$.  Furthermore, we say
that $\alpha$ \emph{fixes a set} $X\subseteq V$ if $\alpha$ fixes
every element of $X$.

\begin{lemma}\label{lem:faith}
If a graph $G$ has a faithful embedding $\Pi$ into some surface $S$,
then $\rig G\le3$.
\end{lemma}

\begin{proof}
  Clearly, $G$ is connected as disconnected graphs don't have a
  cellular embedding.  If $G$ is a path or a cycle, then $\rig G\le2$.
  Otherwise, $G$ contains some vertex $v$ with at least 3 neighbors.
  Notice that a facial walk cannot contain a segment of the form
  $uvu$. Therefore, some facial walk $W$ contains a segment $uvw$,
  where $u$ and $w$ are two different neighbors of $v$. As $v$ has at
  least one further neighbor that is distinct from $u$ and $w$, $uvw$
  cannot be a segment of any other facial walk than $W$.
  
  We now show that $\{u,v,w\}$ is a fixing set.  Assume that $\alpha$
  is an automorphism of $G$ that fixes the   vertices $u$, $v$ and $w$.
  We have to prove that $\alpha$ is the identity.
  
  Note that $v_1v_2\cdots v_k$ is a $\Pi$-facial walk if and only if
  $\alpha(v_1)\alpha(v_2)\cdots\alpha(v_k)$ is a $\Pi^\alpha$-facial
  walk.  Since $\Pi$ and $\Pi^\alpha$ are equivalent and hence, have
  the same facial walks, $\alpha$ takes each $\Pi$-facial walk to a
  $\Pi$-facial walk.  It follows that $\alpha$ takes $W$ onto itself.
  Since $\alpha$ fixes two consecutive vertices of $W$, it actually
  fixes $W$.
  
  Call two $\Pi$-facial walks $W_1$ and $W_2$ \emph{adjacent} if they
  share an edge.  Suppose that adjacent facial walks $W_1$ and $W_2$
  share an edge $u_1u_2$ and that $\alpha$ fixes $W_1$.  Since
  $u_1u_2$ cannot participate in any third facial walk, $\alpha$ takes
  $W_2$ onto itself. Since $u_1$ and $u_2$ are fixed, $\alpha$ fixes
  $W_2$, too.
  
  Now consider the graph whose vertices are the $\Pi$-facial walks
  with the adjacency relation defined as above.  It is not hard to see
  that this graph is connected, implying that $\alpha$ is the identity
  on the whole vertex set $V(G)$.
\end{proof}

\noindent
By the Thomassen Theorem and by Lemma \ref{lem:faith} it follows that
every 3-connected LEW embeddable graph has rigidity index at most 3.
Hence we can apply Theorem \ref{thm:rig} to obtain the following
result.

\begin{corollary}\label{cor:fromTho}
  Let $C$ be any class consisting only of 3-connected LEW embeddable
  graphs.  If $C^*$ has a complete invariant computable in \ac k,
  $k\ge1$, then $C$ has a canonical labeling in~\ac k.
\end{corollary}

\noindent
Noteworthy, the class of all 3-connected LEW embeddable graphs is
recognizable in polynomial time \cite[Theorem 5.1.8]{MT01}.

\begin{lemma}\label{lem:polyhedral}
  If a connected graph $G$ has a polyhedral
  embedding into a surface $S$, then we have $\rig G\le4c$, where $c$
  is the total number of non-equivalent polyhedral embeddings of $G$ into~$S$.
\end{lemma}

\begin{proof}
  To simplify the current exposition, we prove the lemma only for the
  case that $S$ is orientable.  Let $a\in V(G)$ be a vertex in $G$. We
  call two rotation systems $R=\langle G,T\rangle$ and $R'=\langle
  G,T'\rangle$ of $G$ \emph{$a$-coherent} if the binary relations
  $T(a,\cdot,\cdot)$ and $T'(a,\cdot,\cdot)$ coincide.

\begin{claim}
  Let $ab$ be an edge in $G$. Then any isomorphism $\alpha$ between
  two $a$-coherent rotation systems $R=\langle G,T\rangle$ and
  $R'=\langle G,T'\rangle$ of $G$ that fixes both $a$ and $b$, fixes
  also $\Gamma(a)$.
\end{claim}

\begin{proofofclaim}
  Since $\alpha$ fixes $a$, it takes $\Gamma(a)$ onto itself.  Since
  $T(a,\cdot,\cdot)=T'(a,\cdot,\cdot)$, $\alpha$ is an automorphism of
  this binary relation.  The latter is a directed cycle and $\alpha$
  must be a shift thereof.  Since $\alpha$ fixes $b$, it has to fix
  the whole cycle.
\end{proofofclaim}

\noindent
Let $R_1,\ldots,R_{2c}$ (where $R_i=\langle G,T_i\rangle$) be the
rotation systems representing all polyhedral embeddings of $G$ into
$S$ (i.e., each of the $c$ embeddings is represented by two mutually
conjugated rotation systems).  Pick an arbitrary edge $xy$ in $G$.
For each $i$, $1<i\le2c$, select a vertex $x_i$ so that $R_i$ and
$R_1$ are not $x_i$-coherent and the distance between $x$ and $x_i$ is
minimum (it may happen that $x_i=x$). Furthermore, select $y_i$ and
$z_i$ in $\Gamma(x_i)$ so that
$$
T_1(x_i,y_i,z_i)\ne T_i(x_i,y_i,z_i).
$$
We will show that $\{x,y,y_2,z_2,\ldots,y_{2c},z_{2c}\}$ is a fixing set.
Assume that $\alpha\in\aut(G)$ fixes all these vertices.
We have to show that $\alpha$ is the identity.

Notice that $R_1^\alpha$ is a polyhedral embedding of $G$ into $S$
because so is $R_1$. Therefore $R_1^\alpha=R_k$ for some $k\le2c$.
Suppose first that $R_k$ and $R_1$ are $x$-coherent.  We will apply
Claim~1 repeatedly to $R=R_k$ and $R'=R_1$.  We first put $a=x$ and
$b=y$ and see that $\alpha$ fixes $\Gamma(x)$.  If the distance
between $x$ and $x_k$ is more than 1, we apply Claim~1 once again for
$xx'$ being the first edge of a shortest path $P$ from $x$ to $x_k$
(now $a=x'$ and $b=x$; we have $\alpha(x')=x'$ as $x'\in\Gamma(x)$,
and $R_k$ and $R_1$ are $x'$-coherent by our choice of $x_k$).
Applying Claim~1 successively for all edges along $P$ except the last
one, we arrive at the conclusion that $\alpha(x_k)=x_k$. This also
applies for the case that $R_k$ and $R_1$ are not $x$-coherent, when
we have $x_k=x$ by definition.

It follows that $\alpha$ is an isomorphism between the cycles
$T_k(x_k,\cdot,\cdot)$ and $T_1(x_k,\cdot,\cdot)$.  Our choice of
$y_k$ and $z_k$ rules out the possibility that $k\ge2$ and we conclude
that $k=1$. In other words, $R$ and $R'$ are coherent everywhere.
Therefore, we are able to apply Claim~1 along any path starting from
the edge $ab=xy$. Since $G$ is connected, we see that $\alpha$ is the
identity permutation on~$V(G)$.
\end{proof}

\noindent
By the Mohar-Robertson Theorem and Lemma \ref{lem:polyhedral} it
follows that every connected graph having a polyhedral embedding into
a surface $S$ has rigidity index bounded by a constant depending only
on $S$.\footnote{ As we recently learned, this result has been
  independently obtained in \cite{FM} by using a different argument.}
Applying Theorem \ref{thm:rig}, we obtain the following result.

\begin{corollary}\label{cor:fromMRo}
  Let $C$ be any class containing only graphs having a polyhedral
  embedding into a fixed surface $S$. If $C^*$ has a complete
  invariant computable in \ac k, $k\ge1$, then $C$ has a canonical
  labeling in~\ac k.
\end{corollary}

\noindent
We conclude this section by applying a ready-to-use result on the
rigidity index of 5-connected graphs that are embeddable into a fixed
surface $S$.
\bigskip

\noindent
{\bf The Fijav\v{z}-Mohar Theorem.}~\cite{FM} 
{\it The rigidity index of 5-connected graphs embeddable into 
a surface $S$ is bounded by a   constant depending only on $S$.}

\smallskip

\begin{corollary}\label{cor:fromFMo}
  Let $C$ be the class of 5-connected graphs embeddable into a fixed
  surface $S$.  If $C^*$ has a complete invariant computable in \ac k,
  $k\ge1$, then $C$ has a canonical labeling in~\ac k.
\end{corollary}

\section{Conclusion and open problems}\label{s:concl}

\noindent
For several important classes of graphs, we provide \nc{}
Turing-reductions from canonical labeling to computing a complete
invariant. As a consequence, we get a canonical labeling \nc{}
algorithm for graphs with bounded treewidth by using a
known~\cite{GV06} \nc{}-computable complete invariant for such graphs.

We also consider classes of graphs embeddable into a fixed surface.
Though we currently cannot cover this case in full extent, we provide
\nc{} reductions between the canonical labeling and complete invariant
problems for some representative subclasses (namely, 3-connected
graphs with either a polyhedral or an LEW embedding as well as all
embeddable 5-connected graphs).

To the best of our knowledge, complete invariants (even isomorphism
tests) in \nc{} are only known for the sphere but not for any other
surface.  The known isomorphism tests and complete-invariant
algorithms designed in \cite{FM80,Lic80,Mil80,Mil83,Gro00} run in
sequential polynomial time.  Nevertheless, the hypothesis that the
complexity of some of these algorithms can be improved from P to NC
seems rather plausible.  By this reason it would be desirable to
extend the reductions proved in the present paper to the whole class
of graphs embeddable into $S$, for any fixed surface~$S$.  As a first
step in this direction one could consider the class of 4-connected
toroidal graphs.\footnote{As shown in \cite{FM}, graphs in this class
  can have arbitrarily large rigidity index.}

A more ambitious research project is to find an NC-reduction from the
canonical labeling problem to computing a complete invariant for
classes of graphs that are defined by excluding certain graphs as minors or,
equivalently, for classes of graphs closed under minors.  A
polynomial-time canonization algorithm for such classes has been
worked out by Ponomarenko \cite{Pon88}.  Note that any class of graphs
with bounded treewidth as well as any class consisting of all graphs
embeddable into a fixed surface is closed under minors.

\section*{Acknowledgement}
\noindent
We thank Ga\v{s}per Fijav\v{z} and Bojan Mohar for sending us their
manuscript~\cite{FM}.

\end{document}